\documentclass[prl,preprint,twocolumn,10pt]{revtex4-1}

\usepackage{amsmath,amssymb,amsfonts,epsfig,enumerate}

\begin{document}

\preprint{IZTECH-HEP-03/2016}

\title{Curvature-Restored Gauge Invariance and Ultraviolet Naturalness}

\author{Durmu{\c s} Ali Demir}
\email{demir@physics.iztech.edu.tr}
\homepage{http://physics.iyte.edu.tr/staff/durmus-ali-demir/}
\affiliation{Department of Physics, {\.I}zmir Institute of Technology, IZTECH, TR35430, {\.I}zmir, Turkey}

\date{\today}

\begin{abstract}

It is shown that, $(a \Lambda^2 + b |H|^2)R$ in a spacetime of curvature $R$ is a natural ultraviolet $(U\!V)$ completion of $(a \Lambda^4 + b \Lambda^2 |H|^2)$
in the flat-spacetime Standard Model $(S\!M)$ with Higgs field $H$, $U\!V$ scale $\Lambda$ and loop factors $a$, $b$. This curvature completion
rests on the fact that a $\Lambda$-mass gauge theory in flat spacetime turns, on the cut-view $R = 4 \Lambda^2$, into a massless gauge theory in
curved spacetime. It provides a symmetry reason for curved spacetime, wherein gravity and matter are both low-energy effective phenomena.
Gravity arises correctly  if  new physics exists with at least 63 more  bosons than fermions, with no need to interact with the $S\!M$ and
with dark matter as a natural harbinger. It can source various cosmological, astrophysical and collider phenomena depending on its spectrum and
couplings to the $S\!M$.

\end{abstract}

\maketitle

The $S\!M$, spectrally completed with the discovery of its Higgs boson, is experimentally affirmed to describe physics at the Fermi scale, $G_F$.
Its
validity in the $U\!V$ direction comes to an end at a physical scale $\Lambda$ at which it is environed by a new $U\!V$ physics.
Integrating out all trans-Fermi high-frequency fluctuations,  low-energy $S\!M$ fields $\psi_{\!S\!M}$ develop the effective action
\begin{eqnarray}
\label{action-SM-flat}
\!\!\!\!\!\!\!S\!\left(\eta\right) =  S_{G_{F}}\!\!\left({\eta}, \psi_{\!S\!M}, \log\left( G_{F}\Lambda^{2}\right)\right) + S^{0}_{\Lambda}\!\left({\eta}\right) +  S^{1}_{\Lambda}\!\left(\eta\right)
\end{eqnarray}
in the flat spacetime of  metric ${\textstyle{\eta_{\mu\nu}}}$ such that the action
${\textstyle{S_{G_{F}}\!\left({\eta}, \psi_{\!S\!M}, \log\left(G_{F} \Lambda^{2}\right)\!\right)}}$ encodes the tree-level $S\!M$ interactions augmented by logarithmic $U\!V$ contributions,
\begin{eqnarray}
\label{action-UV-0}
S^{0}_{\Lambda}\!\left({\eta}\right) \! =\!\!\! \int\!\! d^4x\! \sqrt{ \left\Vert\eta\right\Vert} \left\{a \Lambda^4\! + a_m  \Lambda^2 m_H^2 + b \Lambda^2 H^{\dagger} H \right\}
\end{eqnarray}
entrains a $U\!V$-sized vacuum energy along with a $U\!V$-sized Higgs boson mass, and
\begin{eqnarray}
\label{action-UV-1}
S^{1}_{\Lambda}\!\left(\eta\right)\! =\!\!\! \int\!\! d^4x\! \sqrt{ \left\Vert\eta\right\Vert} c_V \Lambda^2 \eta_{\mu\nu} {\mbox{Tr}}\!\left\{ V^{\mu} V^{\nu}\right\}
\end{eqnarray}
adds $U\!V$-sized gauge boson masses so that hypercharge, isospin and color are all explicitly broken at the $U\!V$ \cite{peskin}.

In the above, $\Lambda$ is a fundamental scale of nature just as $G_F$. It is a physical scale rather than a formal momentum cutoff introduced to regulate the loop integrals. In this sense, $S\!\left(\eta\right)$ is a physical effective theory whose no part, including $S^{0}_{\Lambda}\!\left({\eta}\right)$ and $S^{1}_{\Lambda}\!\left(\eta\right)$, can be modified at will.  If $\Lambda$ were a formal cutoff it would be possible to eradicate both  $S^{0}_{\Lambda}\!\left({\eta}\right)$ and $S^{1}_{\Lambda}\!\left(\eta\right)$ simply by switching from cutoff regularization to, say, dimensional regularization.

Quarks and leptons, whose masses vary with $\Lambda$ only logarithmically,
stay put at the Fermi scale.  In view of the Higgs mechanism,
therefore,
$S^{0}_{\Lambda}\!\left({\eta}\right)$ and  ${\textstyle{S^{1}_{\Lambda}\!\left(\eta\right)}}$
stand as subversive outliers. They render the $S\!M$ unnatural \cite{natural}. Its natural extensions like supersymmetry, extra dimensions and technicolor have not even glimpsed at collider searches \cite{lhc-search}. It must therefore be naturalized by a different mechanism.
That mechanism, if exists, must be able to {\it (i)} eradicate ${\textstyle{S^{1}_{\Lambda}\!\left(\eta\right)}}$
to restore gauge invariance, {\it (ii)} ameliorate $S^{0}_{\Lambda}\!\left({\eta}\right)$ to
stabilize the Higgs sector, and {\it (iii)} elucidate $\Lambda$ to reveal the $U\!V$
physics behind it. They are approached below by restoring gauge invariance (not by using affine geometry as in \cite{onceki}, though the results agree).

In search for a mechanism to restore gauge symmetries at the $U\!V$, it proves effectual to introduce
\begin{eqnarray}
{{I}}\!\!\left(\eta\right)\!=\!\!\! \int\!\! d^4x\!\sqrt{\! \left\Vert\eta\right\Vert} c_V\! {\mbox{Tr}}\!
\left\{\!\eta_{\mu\nu}\eta_{\alpha\beta} V^{\mu\alpha} V^{\nu\beta}\right\}
\end{eqnarray}
and its by-parts arrangement for $R_{\xi}$ gauge
\begin{eqnarray}
\label{ikiside}
\!\!\!\!\!\!\!{{\tilde{I}}}\!\!\left(\eta\right)\! = \!\!\! \int\!\! d^4x\! \sqrt{\! \left\Vert\eta\right\Vert} 2 c_V {\mbox{Tr}}\! \left\{\!
V^{\mu}\! \left(\! -{\mathcal{D}}^2 \eta_{\mu\nu}\!\! +\! {\mathcal{D}}_{\mu} {\mathcal{D}}_{\nu}\!\! +\!\! V_{\mu\nu}\!\right)\!\!V^{\nu}\!\right\}
\end{eqnarray}
as two gauge kinetic functionals which convert to one another through by-parts integration. This by-parts equivalence of theirs, ${\textstyle{{\tilde{I}}\!\!\left(\eta\right) \equiv {{I}}\!\!\left(\eta\right)}}$,
can be put in use to construct
\begin{eqnarray}
\label{zero-sum-degil}
\!\!\!\!{\tilde{S}}^{1}_{\Lambda}\!\left(\eta\right) = - \frac{1}{2}\!{{I}}\!\left(\eta\right)\! +\! \frac{1}{2}\! {\tilde{I}}\!\left(\eta\right)\!+ \!S^{1}_{\Lambda}\!\!\left(\eta\right)
\end{eqnarray}
as an equivalent of ${\textstyle{S^{1}_{\Lambda}\!\!\left(\eta\right)}}$. Corresponding effective action
\begin{eqnarray}
\!\!\!\!\!\!\!{\tilde{S}}\!\left(\eta\right) =  S_{G_{F}}\!\!\left({\eta}, \psi_{\!S\!M}, \log\left( G_{F}\Lambda^{2}\right)\right) + S^{0}_{\Lambda}\!\left({\eta}\right) +  {\tilde{S}}^{1}_{\Lambda}\!\left(\eta\right)
\end{eqnarray}
is equivalent to $S\!\!\left(\eta\right)$. This dynamical equivalence is, however, quite vulnerable to
spacetime geometry. Indeed, viewing $\eta_{\mu\nu}$
as a specific value assigned to a curved metric $g_{\mu\nu}$ and inductively unassigning it as
\begin{eqnarray}
\label{metric}
\eta_{\mu\nu} \preccurlyeq g_{\mu\nu}
\end{eqnarray}
${\tilde{S}}^{1}_{\Lambda}\!\left(\eta\right)$, as is, can be carried into curved spacetime
to find that ${\tilde{S}}^{1}_{\Lambda}\!\left(g\right) \not\equiv S^{1}_{\Lambda}\!\!\left(g\right)$ simply because
${\textstyle{{\tilde{I}}\!\!\left(g\right) \not\equiv {{I}}\!\!\left(g\right)}}$.
The reason is that ${\textstyle{{\tilde{I}}\!\!\left(g\right)}}$, unlike
${\textstyle{{\tilde{I}}\!\!\left(\eta\right)}}$, must involve not just ${\textstyle{V_{\mu\nu}}}$ but ${\textstyle{V_{\mu\nu} + R_{\mu\nu} (g)}}$, where
${\textstyle{R_{\mu\nu} (g)}}$ is the Ricci curvature  of $g_{\mu\nu}$. This non-equivalence is actually a blessing in disguise for taming the gauge sector.  Indeed, if ${\textstyle{\Lambda^2 g_{\mu\nu}}}$ in ${\textstyle{S^{1}_{\Lambda}\!\!\left(g\right)}}$ is construed as a specific value assigned to  ${\textstyle{R_{\mu\nu}(g)}}$ then its inductive unassignment
\begin{eqnarray}
\label{curvature}
\Lambda^2 g_{\mu\nu} \preccurlyeq R_{\mu\nu}(g)
\end{eqnarray}
takes ${\textstyle{S^{1}_{\Lambda}\!\!\left(g\right)}}$ into
\begin{eqnarray}
\label{action-UV-1-curved}
S^{1}_{\Lambda}\!\left(R\right)\! =\!\! \int \!\! d^4x\! \sqrt{ \left\Vert g\right\Vert}  c_V R_{\mu\nu}(g) {\mbox{Tr}}\!\left\{ V^{\mu} V^{\nu}\right\}
\end{eqnarray}
with which ${\textstyle{S^{1}_{\Lambda}\!\!\left(R\right) + (1/2) {\tilde{I}}\!\!\left(g\right)  = (1/2) {{I}}\!\!\left(g\right)}}$ and thus
\begin{eqnarray}
\label{zero-sum}
{\tilde{S}}^{1}_{\Lambda}\!\left(R\right) = - \frac{1}{2} {{I}}\!\left(g\right) + \frac{1}{2} {\tilde{I}}\!\left(g\right)+ S^{1}_{\Lambda}\!\left(R\right)  \equiv  0
\end{eqnarray}
so that all gauge symmetries are restored at the $U\!V$. The unassignment of $g_{\mu\nu}$ ($R_{\mu\nu}$) renders the Poincare (gauge) breaking by $\Lambda$ futile.
The classical curvature restores gauge invariance to naturalize the $S\!M$ gauge sector.

This naturalization mechanism extends over entirety of the $S\!M$ if ${\tilde{S}}\!\left(\eta\right)$ is construed as a curved spacetime
effective action ${\tilde{S}}\!\left(R\right)$ evaluated at the  curvature  $R_{\mu\nu}(g)
= \Lambda^2 g_{\mu\nu}$ and then at the metric $g_{\mu\nu} = \eta_{\mu\nu}$. Namely, ${\tilde{S}}\!\left(\eta\right)$ is
nothing but the $U\!V$-scale cut-view of ${\tilde{S}}\!\left(R\right)$
\begin{eqnarray}
\label{renorm}
{\tilde{{S}}}\!\left(\eta\right) = \left. {\tilde{{S}}}\!\left(R\right)\right\vert_{\begin{array}{l}
{\text{set first}}\\
\!{R_{\mu\nu}(g) = \Lambda^2 g_{\mu\nu}}\\
{\text{and then}}\\
{g_{\mu\nu} = \eta_{\mu\nu}}
\end{array}}
\end{eqnarray}
such that ${\tilde{S}}\!\left(\eta\right)$ and ${\tilde{S}}\!\left(R\right)$ both
embrace $\Lambda$, albeit with different physical meanings. Physically, ${\tilde{S}}\!\left(R\right)$ must involve
\begin{enumerate}[(a)]
\item no extra couplings not found in ${\tilde{S}}\!\left(\eta\right)$ as no quantum fluctuations
are left to induce any new coupling,

\item no extra forces except gravity as spacetime can attain required elasticity
if $\Lambda$ nears  gravitational scale \cite{sakharov},
\end{enumerate}
so that it comes to take the familiar Einstein-Hilbert form
\begin{eqnarray}
\label{action-UV-0-curved}
\!\!\!\!\!\!\!\!\!\!\!{\tilde{S}}\!\left(R\right) \!&=& \!S_{G_{F}}\!\!\left(g, \psi_{\!S\!M}, \log\left( G_{F}\Lambda^{2}\right)\right)\nonumber\\
\!&+&\!\!\!\int\!\!\!\!\ d^4x\! \sqrt{ \left\Vert g \right\Vert} \! \left( {a} \Lambda^2 + {a_m} m_H^2  + {b}  H^{\dagger} H \right)\! \frac{{R}(g)}{4}
\end{eqnarray}
though already at one loop (see \cite{onceki} for a variant study)
\begin{eqnarray}
a = \frac{1}{64 \pi^2} (n_b-n_f)
\end{eqnarray}
is negative for $n_b = 28$ bosonic and $n_f = 90$ fermionic degrees of freedom in the $S\!M$. This means that gravity can be induced properly only if the $S\!M$ is extended by
\begin{eqnarray}
\label{sart}
{{n}}^{{{N\!P}}}_b - {{n}}^{{{N\!P}}}_f \geq 63
\end{eqnarray}
new fields belonging to some `new physics' ($N\!P$) sector lying at a scale ${\textstyle{G_{\!N\!\!\!P}}}$. Its effective action is of the form
\begin{eqnarray}
\label{action-SM-flat-NP}
\!\!\!\!\!\!\!\!\!{\tilde{S}}^{N\!P}\!\!\!\left(\eta\right)\! =\!\! S_{G_{\!N\!\!\!P}}\!\!\left(\!{\eta}, \psi_{\!N\!P},\!\log\!\left(\!G_{\!N\!\!\!P}\!\Lambda^{2}\right)\!\right)\!+\! S^{0, N\!P}_{\Lambda}\!\!\!\left({\eta}\right) \!+\!  {\tilde{S}}^{1, N\!P}_{\Lambda}\!\!\!\left(\eta\right)
\end{eqnarray}
if it is secluded from the $S\!M$. If it has no scalars \cite{natural}
\begin{eqnarray}
\label{action-UV-NP}
S^{0, N\!P}_{\Lambda}\!\left({\eta}\right) = \int d^4 x \sqrt{ \left\Vert\eta\right\Vert}\, {{a}^{{{N\!P}}}} \Lambda^{4}
\end{eqnarray}
so that, through (\ref{renorm}), ${\tilde{S}}\!\left(\eta\right)$ and ${\tilde{S}}^{N\!P}\!\!\left(\eta\right)$ add up to give
\begin{eqnarray}
\label{action-Fermi-curve}
\!\!\!\!\!\!\!\!{\tilde{S}}_{S\!M\!+\!N\!P}\!\left(R\right) &=&
{{S}}_{G_{F}}\!\!\left({g}, \psi_{\!S\!M}, \log\left( G_{F}\Lambda^{2}\right)\right)\nonumber\\
&+& S_{G_{\!N\!\!\!P}}\!\!\left({g}, \psi_{\!N\!P}, \log\left(G_{\!N\!\!\!P} \Lambda^{2}\right)\right)\\
&+& \int\!\! d^{4}x \sqrt{\Vert g \Vert}\! \left( \frac{R\left(g\right)}{16 \pi G_N} + \zeta_{H} R\left(g\right) H^{\dagger} H  \!\right)\nonumber
\end{eqnarray}
as a completely $U\!V$-natural effective field theory governing $\psi_{\!S\!M}$ and $\psi_{\!N\!P}$ dynamics in curved spacetime \cite{onceki} such
that  $\zeta_H = b/4$ is Higgs-curvature direct coupling, and
\begin{eqnarray}
\label{params}
G_N = \left(4 \pi \left(a + {a}^{{{N\!P}}}\right) \Lambda^2 +  4 \pi a_m m_H^2 \right)^{-1}
\end{eqnarray}
is Newton's  constant (see also \cite{demir2}). Not only these curvature couplings but also those in $S_{G_{F}}\!\!\left({g}, \psi_{\!S\!M}, \log\left(\! G_{F}\Lambda^{2}\!\right)\right)$
and $S_{G_{\!N\!\!\!P}}\!\!\left({g}, \psi_{\!N\!P}, \log\left(G_{\!N\!\!\!P}\Lambda^{2}\!\right)\right)$ are all
given by couplings in the flat spacetime effective actions ${\tilde{S}}\!\left(\eta\right)$ and ${\tilde{S}}^{N\!P}\!\!\left(\eta\right)$.

In (\ref{action-Fermi-curve}), vacuum energy, already naturalized in the $U\!V$ with corrections $\propto (m_H^2)^2\! \log\!\left(G_{F}\Lambda^{2}\right)$,
needs be naturalized also in the $I\!R$ as it is still far bigger than the observational
value of $m_{\nu}^4$. This is the $I\!R$ cosmological constant problem \cite{demir2}.

In (\ref{action-Fermi-curve}), the $N\!P$, secluded from the $S\!M$, is a natural home to noninteracting dark matter. Its spectrum, modelable as
$SU(6)$ gauge theory or $SU(7)$ with two fermions or
anything with ${{n}}^{{{N\!P}}}_b - {{n}}^{{{N\!P}}}_f \geq 63$,
is secluded enough to form a dark matter observable via only its weight \cite{dm}.

In (\ref{action-Fermi-curve}), there exist no higher-curvature terms
because, after inducing the Einstein-Hilbert term,  no suitable couplings
are left in the $S\!M$.
If needed, they can be added by hand but that
reduces predictive power of the $S\!M\!+\!N\!P$.
Indeed, adding, for instance, the ghost-free quadratic action $\int \!\! d^{4}x \sqrt{\Vert g \Vert} \left\{
\omega W^2(g) \!+\! \gamma \left( R(g)^2 \!-\! 6 G^2(g)\right)\! \right\}$
does not unnaturalize (\ref{action-Fermi-curve}) yet ambiguates it by the undetermined
coefficients  of Weyl $(\omega)$ and Gauss-Bonnet $(\gamma)$ invariants.

In (\ref{action-Fermi-curve}), logarithmic $U\!V$ contributions coming through $\log\left(\! G_{F}\Lambda^{2}\right)$
lead to multiplicative renormalizations of  the ${{S}}_{G_{F}}\!\!\left({g}, \psi_{\!S\!M}, \log\left( G_{F}\Lambda^{2}\right)\right)$ and
$S_{G_{\!N\!\!\!P}}\!\!\left({g}, \psi_{\!N\!P}, \log\left(G_{\!N\!\!\!P} \Lambda^{2}\right)\right)$.
These logarithms can actually be construed as loop integrals in a $D=4-\epsilon$ dimensional momentum space of total volume
$\mu^{2 \epsilon} \infty^{4- 2\epsilon}$ so that the formal equivalence
$\log\!\left( G_{F}\Lambda^{2}\right) \equiv {2}/{\epsilon} + \log G_F \mu^2$
with small but finite $\epsilon$ enables the two logarithmic parts of (\ref{action-Fermi-curve}) to be formulated in the dimensional regularization scheme with associated
renormalization methods \cite{peskin}.

In (\ref{action-Fermi-curve}), the $S\!M$ is decoupled from the $N\!P$. Their coupling can cause new effects. One possibility is Higgs  mass shifts like ${\textstyle{G_{\!N\!\!\!P}^{-1/2}\! \log\left(G_{\!N\!\!\!P}\Lambda^2\!\right)}}$,
which beget a  logarithmic unnaturalness as in supersymmetry unless ${\textstyle{G_{\!N\!\!\!P} \cong G_{\!F}}}$ \cite{martin}. The interacting dark matter under search in direct detection experiments is another possibility. The recent LHC diphoton signal \cite{diphoton} can well be a precursor of such an $N\!P$.

In (\ref{action-Fermi-curve}), matter and gravity meet in a physically consistent framework in that they  both are sub-Fermi effective interactions. This is a crucial property because putting quantized matter into curved geometry is an aporia as quantum gravity is distant and classical gravity is inconsistent \cite{seif}. They do not 
enhance naturalness \cite{natural2}.

In summary, restoration of gauge invariance has led to naturalization of the $S\!M$ via gravity. This is
not surprising because, at least in macroscopic world, it is gravity that dictates what is natural and what is unnatural. It
causes the Pisa tower to be sensed as unnatural. Holding a pen upright on its tip, while impossible in a plane without two counter-balancing forces
(Higgs coupling to top quark and  top squark), is possible in an inclined plane with a single force
(Higgs coupling to top quark) \cite{talk}. It is by imaginations as such that unnatural
${\tilde{S}}\!\left(\eta\right) + {\tilde{S}}^{N\!P}\!\!\left(\eta\right)$ in flat spacetime turns into natural ${\tilde{S}}_{S\!M\!+\!N\!P}\!\left(R\right)$ in curved spacetime.
The mechanism makes, as were also with the previous work \cite{onceki}, three salient predictions:
\begin{enumerate}[(I)]
\item Gravity arises as a large-distance effective force consistently coupled to
the low-energy quantum effective action. It is the requisite $U\!V$ physics that completes the $S\!M$ and renders  the $\Lambda$ physical.

\item New physics exists as a highly crowded sector ($\Lambda < (8 \pi G_N)^{-1/2}$ for ${{n}}^{{{N\!P}}}_b - {{n}}^{{{N\!P}}}_f > 128 \pi^2 + 62 \approx 1325$)
which does not have to couple to the $S\!M$ matter. This secluded sector, which can in principle lie at any scale sufficiently below $\Lambda$,  can source  dark matter as a non-interacting non-baryonic matter which can be sensed via only its weight. In this setup, the ${S\!M\!+\!N\!P}$ is all $U\!V$-natural.

\item New physics may interact with the $S\!M$ partly or wholly. In this case,  scalar fields in the $N\!P$, even its vector-like fermions, can cause the Higgs boson mass to shift by  ${\textstyle{G_{\!N\!\!\!P}^{-1/2}}}$ depending on coupling strengths. They destabilize the $S\!M$  unless ${\textstyle{G_{\!N\!\!\!P} \cong G_{\!F}}}$. The $L\!H\!C$ diphoton signal, if real, may be stemming from such an $N\!P$ sector.
\end{enumerate}
Future research will reveal more about the formalism.

This work is supported in part by the T{\"U}B{\.I}TAK grant 115F212. It is dedicated to honorable memory of Prof. Nam{\i}k K. Pak, who was a dedicated teacher, an indefatigable researcher and a great friend.

\end{document}